\documentclass[8.5pt,twoside,twocolumn]{article}
\usepackage[super,sort&compress,comma]{natbib} 
\usepackage[version=3]{mhchem}
\usepackage{balance}
\usepackage{times,mathptm}
\usepackage{sectsty}
\usepackage{graphicx} 
\usepackage{lastpage}
\usepackage[format=plain,justification=raggedright,singlelinecheck=false,font=small,labelfont=bf,labelsep=space]{caption} 
\usepackage{fancyhdr}
\begin{document}

\thispagestyle{plain}
\fancypagestyle{plain}{
\fancyhead[L]{\includegraphics[height=8pt]{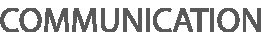} }
\fancyhead[C]{\hspace{-1cm}\includegraphics[height=20pt]{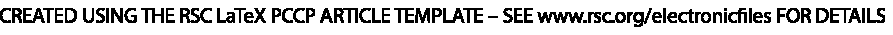}}
\renewcommand{\headrulewidth}{1pt}}
\renewcommand{\thefootnote}{\fnsymbol{footnote}}
\renewcommand\footnoterule{\vspace*{1pt}%
\hrule width 3.4in height 0.4pt \vspace*{5pt}} 
\setcounter{secnumdepth}{5}

\makeatletter 
\renewcommand\@biblabel[1]{#1}            
\renewcommand\@makefntext[1]%
{\noindent\makebox[0pt][r]{\@thefnmark\,}#1}
\makeatother 
\renewcommand{\figurename}{\small{Fig.}~}
\sectionfont{\large}
\subsectionfont{\normalsize} 

\fancyfoot{}
\fancyfoot[LO,RE]{\vspace{-7pt}\includegraphics[height=9pt]{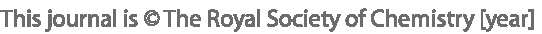}}
\fancyfoot[CO]{\vspace{-7.2pt}\hspace{12.2cm}\\[-5pt]{\scriptsize{\textsf{Typeset: \today}}}}
\fancyfoot[CE]{\vspace{-7.5pt}\hspace{-13.5cm}\\[-5pt]{\scriptsize{\textsf{Typeset: \today}}}}
\fancyfoot[RO]{\footnotesize{\sffamily{1--\pageref{LastPage} ~\textbar  \hspace{2pt}\thepage}}}
\fancyfoot[LE]{\footnotesize{\sffamily{\thepage~\textbar 1--\pageref{LastPage}}}}
\fancyhead{}
\renewcommand{\headrulewidth}{1pt} 
\renewcommand{\footrulewidth}{1pt}
\setlength{\arrayrulewidth}{1pt}
\setlength{\columnsep}{6.5mm}
\setlength\bibsep{1pt}

\twocolumn[
  \begin{@twocolumnfalse}
\noindent\LARGE{\textbf{Negative area compressibility in silver(I) tricyanomethanide$^\dag$}}
\vspace{0.6cm}

\noindent\large{\textbf{Sarah A.\ Hodgson,$^{\textrm a}$ Jasper Adamson,$^{\textrm a}$ Sarah J. Hunt,$^{\textrm a}$ Matthew J. Cliffe,$^{\textrm a}$ Andrew B.\ Cairns,$^{\textrm a}$\\ Amber L. Thompson,$^{\textrm a}$ Matthew G. Tucker,$^{\textrm{b,c}}$ Nicholas P. Funnell,$^{\textrm a}$  and Andrew L.\ Goodwin$^{\textrm a,\ast}$}}\vspace{0.5cm}

\noindent\textit{\small{\textbf{Received Xth XXXXXXXXXX 20XX, Accepted Xth XXXXXXXXX 20XX\newline
First published on the web Xth XXXXXXXXXX 200X}}}

\noindent \textbf{\small{DOI: 10.1039/b000000x}}
 \end{@twocolumnfalse} \vspace{0.6cm}
  ]

\noindent\textbf{The molecular framework Ag(\textit{\textbf tcm}) (\textit{\textbf tcm}$^{\boldsymbol -}$ = tricyanomethanide) expands continuously in two orthogonal directions under hydrostatic compression. The first of its kind, this negative area compressibility behaviour arises from the flattening of honeycomb-like layers during rapid pressure-driven collapse of the interlayer separation.}
\section*{}
\vspace{-1cm}
\footnotetext{\dag~Electronic Supplementary Information (ESI) available: Synthesis; experimental methods; X-ray single-crystal diffraction refinement details (variable-temperature and variable-pressure); neutron powder diffraction refinement details; calculations.}
\footnotetext{$^{\textrm a}$ Department of Chemistry, University of Oxford, Inorganic Chemistry Laboratory, South Parks Road, Oxford OX1 3QR, U.K. Fax: +44 1865 274690; Tel: +44 1865 272137; E-mail: andrew.goodwin@chem.ox.ac.uk. $^{\textrm b}$ ISIS Facility, Rutherford Appleton Laboratory, Chilton, Didcot, Oxfordshire, OX11 0QX, U.K. $^{\textrm c}$ Diamond Light Source, Harwell Campus, Didcot, Oxfordshire, OX11 0DE, U.K.}

In the absence of a structural phase transition, it is a thermodynamic requirement that a system reduce its volume under hydrostatic pressure.\cite{Nicolaou_2012} The phenomenon of negative compressibility is the rare and counterintuitive effect whereby this volume reduction couples to an expansion of the material in at least one linear dimension.\cite{Baughman_1998} Not only does the existence of such behaviour challenge our understanding of ways in which materials can respond to external stimuli, but negative compressibility is also of practical importance. It promises a means of developing artificial muscles, nanoscale actuators and high-performance pressure sensors for sonar and altitude measurements.\cite{Baughman_1998,Aliev_2009} In principle, the most extreme such response allowed thermodynamically is that of negative area compressibility (NAC), whereby a material expands along two orthogonal directions on increasing pressure. To the best of our knowledge, there are no materials known to exhibit an intrinsic NAC response, even if calculations suggest they should exist:\cite{Seyidov_2010} instead the remarkably few negative compressibility materials that have been characterised experimentally expand along only one principal axis when compressed hydrostatically (\emph{i.e.}, negative linear compressibility, NLC).\cite{Goodwin_2008c,Fortes_2011,Li_2012,Shepherd_2012,Cairns_2013} The distinction between NLC and NAC is more fundamental than magnitude alone: NAC materials increase their surface area under hydrostatic pressure, and so they can be used as substrates to provide order-of-magnitude amplification of piezoelectric response in \emph{e.g.}\ ferroelectric sensors.\cite{Baughman_1998}

\begin{figure}[t]
\centering
  \includegraphics{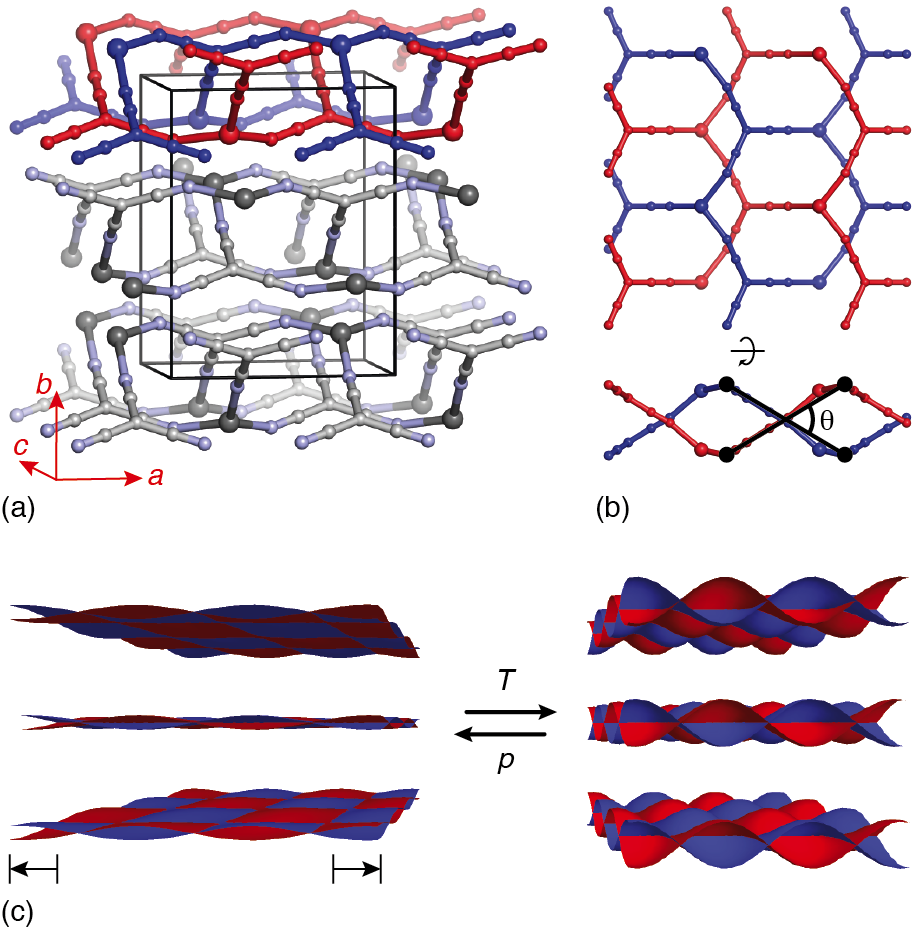}
  \caption{(a) The crystal structure of Ag({\it tcm}) is composed of stacked, doubly-interpenetrating (6,3)-nets in which both Ag$^+$ (large spheres) and {\it tcm}$^-$ ions (smaller spheres) function as trigonal nodes. (b) The extent of rippling of layers (viewed along $\mathbf b$ and $\mathbf a$ in the top and bottom parts of this panel, respectively) can be quantified by the inter-network angle $\theta$ (see SI). (c) The generic mechanical response of a layered material links volume expansion to increasing interlayer separation and increasing layer rippling. The cross-sectional area of each layer is maximised by reducing $\theta$, and hence increases during volume reduction under hydrostatic compression.\label{fig1}}
\end{figure}

One strategy for identifying negative compressibility candidates is to target framework materials that exhibit anisotropic negative thermal expansion (NTE): a material that shrinks in one direction on heating (where volume expansion is likely, if not absolutely required) is likely to expand in that same direction on compression.\cite{Ogborn_2012} While there is no strict requirement that deformation mechanisms under temperature and pressure be identical,\cite{Munn_1972} there is considerable empirical evidence of such a correspondence amongst molecular framework materials.\cite{Goodwin_2008c,Fortes_2011,Cairns_2012,Cairns_2013} Consequently, in order to identify NAC candidates, the natural strategy is to investigate framework materials that exhibit area-NTE.

It was in this context that we chose to explore the mechanical behaviour of silver(I) tricyanomethanide, Ag({\it tcm}): its layer-like network topology [Fig.~\ref{fig1}(a)] is of the general form often associated with area-NTE.\cite{Lifshits_1952,Cliffe_2012} In this compound, each Ag$^+$ cation is coordinated by three {\it tcm}$^-$ anions in an approximately trigonal arrangement; likewise each {\it tcm}$^-$ anion is coordinated by three Ag$^+$ centres.\cite{Konnert_1966,Batten_1998} The resulting hexagonal (6,3) topology is sufficiently open that two honeycomb networks interpenetrate within each layer of the crystal structure [Fig.~\ref{fig1}(b)]. Adjacent layers interact \emph{via} long Ag$\cdots$N contacts ($d$(Ag$\cdots$N)$>$3\,\AA) that are sufficiently weak that the material behaves essentially as a two-dimensional framework, capable even of intercalation chemistry.\cite{Batten_1998}

\begin{figure}[t]
\centering
\includegraphics{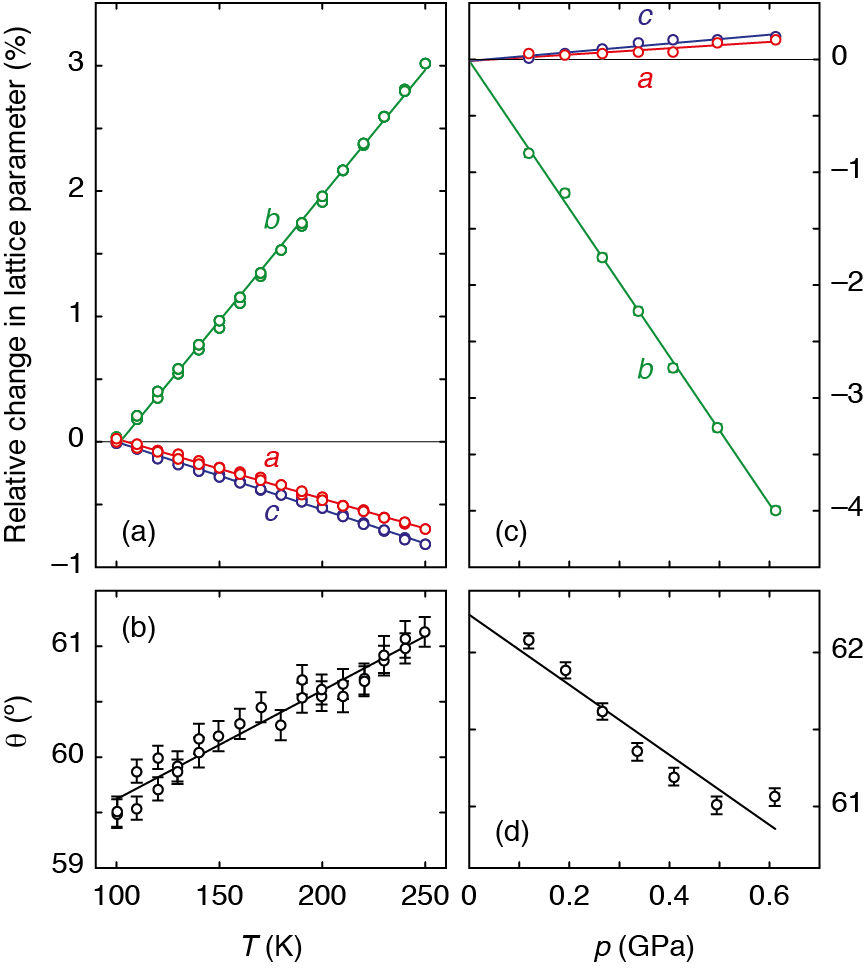}
  \caption{(a) The thermal variation of lattice parameters, determined using single-crystal X-ray diffraction measurements, is illustrated here for ease of comparison as a \% change relative to the extrapolated value at 100\,K. Solid lines are weighted linear fits to the data obtained using the program {\sc pasc}al.\cite{Cliffe_2012} (b) The corresponding variation in inter-network angle $\theta$ indicates an increase in layer rippling on heating. (c) The pressure dependence of lattice parameters, measured using variable-pressure neutron diffraction, is represented as a \% change relative to the extrapolated value at 0\,GPa. (d) The corresponding variation in $\theta$ now indicates a decrease in layer rippling under compression.\label{fig2}}
\end{figure}

Using single-crystal X-ray diffraction measurements, we first established the thermal expansion behaviour of Ag({\it tcm}) in order to verify whether or not the material exhibitx area-NTE. Because the material crystallises in the orthorhombic (polar) space group $Ima2$,\cite{Konnert_1966} its thermal expansivity is uniquely determined by the relative rate of change of the unit cell lengths with respect to temperature.\cite{Cliffe_2012} We observe essentially linear thermal variation in these parameters over the temperature range 100--250\,K [Fig.~\ref{fig2}(a)], yielding coefficients of thermal expansion of $-48(3)$, $200(2)$ and $-54.0(4)$\,MK$^{-1}$ along the $a$, $b$, and $c$ axes, respectively.\cite{Cliffe_2012} As is increasingly found to be the case for molecular frameworks,\cite{Goodwin_2008b,Yang_2009,Ogborn_2012} the magnitudes of these values are large indeed compared to those obtained for traditional engineering materials, for which one expects linear coefficients of thermal expansion in the vicinity of 20\,MK$^{-1}$ (\emph{i.e.}, a 0.2\% increase in linear dimension for each 100\,K temperature rise).\cite{Newnham_2005} What is clear is that Ag({\it tcm}) does indeed exhibit area-NTE, with an area coefficient of thermal expansion for the $(a,c)$-plane, $\alpha_{(010)}=-102(3)$\,MK$^{-1}$, that is the most extreme negative value reported for a layered material. The material undergoes a 1\% reduction in area for each 100\,K rise in temperature---more than an order of magnitude larger than the corresponding values for area-NTE systems such as graphite and Ni(CN)$_2$ ($\alpha_A=-2.4$ and $-13$\,MK$^{-1}$, respectively).\cite{Cliffe_2012,Bailey_1970,Hibble_2007}

The microscopic mechanism responsible for this behaviour can be deduced from the variation in framework geometry observed crystallographically during heating. The standard model of area-NTE in layered materials predicts coupling between NTE and the extent of layer `rippling', as quantified by the inter-network torsion angle $\theta$ [Fig.~\ref{fig1}(b,c)].\cite{Lifshits_1952} Our crystallographic measurements do indeed indicate an increase in $\theta$ with temperature [Fig.~\ref{fig2}(b)]. Moreover, using simple geometric arguments (see SI), the relative rate of change of $\theta$ with respect to temperature can be used to estimate the value of $\alpha_A$ attributable to a layer-rippling mechanism:
\begin{equation}
\alpha_A^{\textrm{calc}}=-\theta\tan\left(\frac{\theta}{2}\right)\alpha_\theta,\label{thetaeq}
\end{equation}
where $\alpha_\theta={\rm d}(\ln\theta)/{\rm d}T$ (see SI for derivation). Our data give $\alpha_A^{\textrm{calc}}=-101$\,MK$^{-1}$, which is within experimental error of the observed value of $-102(3)$\,MK$^{-1}$. Consequently, we can be confident that the mechanism illustrated in Fig.~\ref{fig1}(c) is responsible for area-NTE in Ag({\it tcm}).

In order to determine whether or not Ag({\it tcm}) exhibits the target property of NAC, we performed a series of neutron diffraction measurements across the pressure range $0<p<2$\,GPa using the PEARL instrument at ISIS. The use of neutron radiation helps maximise the scattering contrast between Ag and C/N atoms, which is all the more important for measurements performed in the high-background sample environments required for variable-pressure experiments. We found that, for powdered samples, the ambient phase is stable up to 0.615(6)\,GPa, but by 0.668(6)\,GPa it has transformed reversibly to a related phase with monoclinic symmetry. The same transformation occurs at higher pressures ($>0.8$\,GPa) in single-crystal experiments (see SI for further discussion); our focus here is on the pressure-dependent behaviour of the ambient phase rather than the nature of the phase transition or the high-pressure phase. Rietveld refinement against the corresponding neutron diffraction patterns allowed us to determine the pressure-dependence of the unit cell dimensions throughout the stability field of the ambient phase. The anticipated increase in $a$ and $c$ parameters is evident from these data [Fig.~\ref{fig2}(c)], with the anomalous expansion of the material in these directions being allowed by a much more rapid decrease in the interlayer spacing (given by the lattice parameter $b$). Our variable-pressure lattice parameter data can be converted into compressibilities using the relationship $K_\ell=-\left[\partial(\ln\ell)/{\partial p}\right]_T$, from which we obtain $K_a=-3.5(6)$\,TPa$^{-1}$, $K_b=+66(20)$\,TPa$^{-1}$, and $K_c=-4.0(6)$\,TPa$^{-1}$. Consequently, the compressibility is negative everywhere within the $(a,c)$-plane, with the magnitude of NAC quantified by the corresponding area compressibility $K_{(010)}=K_a+K_c=-7.5(8)$\,TPa$^{-1}$.

Traditional engineering materials (\emph{e.g.}\ iron) have linear compressibilities $K\sim5$\,TPa$^{-1}$ that correspond to a decrease of \emph{ca} 0.5\% in length for each 1\,GPa pressure increase;\cite{Zhang_2010c} consequently, the NAC behaviour of Ag({\it tcm}) is essentially as strong as the positive area compressibility of these `normal' phases. Moreover, despite the fact that the orthorhombic symmetry of Ag({\it tcm}) allows for different behaviour in the $\mathbf a$ and $\mathbf c$ directions, we find that both its compressibility and its thermal expansivity are essentially isotropic within the entire $(a,c)$ plane. Isotropy is important for NAC applications because it avoids composite warping and performance breakdown.\cite{Baughman_1998}

That NAC is driven by the same layer-rippling mechanism as area-NTE is evident in the variation of the parameter $\theta$ with increasing pressure [Fig.~\ref{fig2}(d)]. The volume reduction accommodated by rapid collapse of the interlayer spacing (note $K_b\gg5$\,TPa$^{-1}$) evidently couples with a reduction in the degree of layer rippling: ${\rm d}\theta/{\rm d}p<0$. To substantiate further our interpretation of the NAC mechanism, we recast equation \eqref{thetaeq} in terms of compressibilities; our refined values of $\theta$ give $K_A^{\textrm{calc}}=-\theta\tan(\theta/2)K_\theta=-26$\,TPa$^{-1}$. That the measured area compressibility $K_{(010)}=-7.5(8)$\,TPa$^{-1}$ is somewhat smaller in magnitude reflects the expected $\sim$10\,TPa$^{-1}$ contribution from (positive) bond compressibilities omitted from the geometric formalism of Eq.~\eqref{thetaeq}, as discussed elsewhere for NLC materials.\cite{Ogborn_2012} Indeed, that the $p/T$ correspondence inferred in Fig.~\ref{fig1}(c) holds qualitatively for Ag({\it tcm}) is immediately evident in the close relationship between the temperature- and pressure-induced lattice parameter and layer rippling parameter variations documented in Fig.~\ref{fig2}.

In conclusion, we have demonstrated that silver(I) tricyanomethanide exhibits NAC over the pressure range $0<p<0.612$\,GPa with an area compressibility of similar magnitude (but opposite sign) to that of established engineering materials. Despite the orthorhombic crystal symmetry of this system, this NAC behaviour is essentially isotropic. The phenomenon can be understood in terms of pressure-driven damping of layer `rippling', acting to increase the layer cross-sectional area at larger hydrostatic pressures. In favour of immediate application of Ag({\it tcm}) in ultra-high-precision sensing devices are (i) the ease with which large single crystal samples can be prepared under ambient conditions and using readily-available solvents, and (ii) its optical transparency. Moreover, the polarity of its crystal structure offers the possibility of designing multifunctional devices based on Ag({\it tcm}) in which the interplay between responses to electric field, temperature, and pressure are exploited to produce next-generation piezoelectric and thermoelectric devices.

The authors gratefully acknowledge financial support from the E.P.S.R.C. (EP/G004528/2) and the E.R.C. (Grant Ref.: 279705), the provision of neutron beamtime on the PEARL instrument at ISIS, and the assistance of Dr Marek Jura (ISIS).


\footnotesize{
\bibliography{cc_2013_agtcm} 
\bibliographystyle{rsc} }

\end{document}